\documentclass[twocolumn,showpacs,amsmath,amssymb,superscriptaddress,prl]{revtex4}

\usepackage{graphicx}
\usepackage{dcolumn}
\usepackage{bm}
\usepackage{epsf}
\usepackage{amsmath}
\usepackage{amssymb}
\usepackage{color}

\newcommand{\bra}[1]{\langle #1|}
\newcommand{\ket}[1]{|#1\rangle}

\newcommand{\tr}[1]{\mathrm{tr}\left\{#1\right\}}

\newcommand{\la}{\left\langle}
\newcommand{\ra}{\right\rangle}

\newcommand{\td}{\mathrm{d}}
\newcommand{\ma}[1]{\max{\left\{#1\right\}}}

\newcommand{\co}[1]{\cos{\left(#1\right)}}
\newcommand{\si}[1]{\sin{\left(#1\right)}}

\newcommand{\bla}{bla\\bla\\bla\\bla\\bla}

\newcommand{\PRA}{Phys. Rev. A }

\newcommand{\PRL}{Phys. Rev. Lett. }

\newcommand{\RMP}{Rev. Mod. Phys. }

\newcommand{\mc}[1]{\mathcal{#1}}

\newcommand{\mrm}[1]{\mathrm{#1}}

\begin{document}

\title{Quantum speed limit for non-Markovian dynamics}
\author{Sebastian Deffner}
\affiliation{Department of Chemistry and Biochemistry and Institute for Physical Science and Technology, University of Maryland, 
College Park, Maryland 20742, USA}

\author{Eric Lutz}
\affiliation{Dahlem Center for Complex Quantum Systems, FU Berlin, D-14195 Berlin, Germany}

\date{\today}

\begin{abstract}
We derive a Margolus-Levitin type bound on the minimal  evolution time of an arbitrarily driven open quantum system. We express this quantum speed limit time in terms of the operator norm of the nonunitary generator of the   dynamics. We apply these results to the damped Jaynes-Cummings model and demonstrate that the corresponding bound is tight. We further show that non-Markovian effects can speed up quantum evolution and therefore lead to a smaller quantum speed limit time.\end{abstract}
\maketitle

What is the maximal speed of evolution of a quantum system? This question is of fundamental importance in virtually all areas of  quantum physics, ranging from quantum communication \cite{bek81}, computation \cite{llo00} and metrology \cite{gio11} to  optimal control theory \cite{can09} and nonequilibrium thermodynamics \cite{def10}. For closed systems, the quantum speed limit time which  determines the maximum rate of evolution can  be obtained by combining the results of Mandelstam-Tamm (MT) \cite{man45} and Margolus-Levitin (ML) \cite{mar98}: it is given  by $\tau_\mrm{QSL}= \mbox{max}\{\pi\hbar/(2\Delta E),\pi \hbar/(2E)\}$, where $\Delta E$ is the variance of the energy of the initial state and $E$ its mean energy with respect to the ground state. The minimal time a quantum system needs to evolve from a given state to an orthogonal state is thus determined by its initial energy \cite{fle73,bha83,ana90,vai91,uff93,bro03}. The latter reflects the fact that the Hamiltonian is the generator of  unitary Schr\"odinger dynamics. It is worth emphasizing that the existence of a speed limit time is a purely quantum effect which  vanishes when $\hbar$ goes to zero. Generalizations of the MT and ML findings to  nonorthogonal states and to driven  systems have been provided in Refs.~\cite{gio03,jon10,zwi12} and \cite{pfe93,pfe95,def11}. Recently, the quantum speed limit time has been derived for open systems described by  positive nonunitary maps; applications to dephasing in noisy channels  and  quantum parameter estimation have been discussed \cite{tad13,cam13}. In both approaches, the speed limit time was obtained  in terms of the variance of the generator of the evolution, which reduces to the MT expression in the case of closed system dynamics. To our knowledge, no  ML type  bound has been proposed for open quantum systems to date. 

In this paper, we use a geometric approach to derive a quantum speed limit time   valid for  open system dynamics with possibly time-dependent nonunitary generators. In contrast to previous studies, we obtain a bound that depends on the mean of the  generator; it therefore reduces to the ML formula for unitary processes. MT type bounds are usually derived with the help of the Cauchy-Schwarz inequality. The latter invariably leads to expressions containing the variance of the generator, and is hence not suitable for getting ML type bounds. We here solve this technical challenge by making use of the von Neumann trace equality \cite{neu37,mir75,gri91}. In the following, we  obtain a  quantum speed limit time that depends on  the operator norm of the nonunitary generator and show that the so obtained ML  bound is not only sharper than the MT  bound, it is also tight. By employing both inequalities, we are  able to obtain a unified quantum speed limit time for generic positive open system dynamics. We further apply these results to investigate the influence of non-Markovianity  on the rate of quantum evolution. Non-Markovian (or memory)  effects become important when the relaxation time of the system is comparable to the relaxation time of the environment \cite{bre12,bre07}. They have been shown to play a central role in the creation of steady state entanglement \cite{hue12} and in the description of quantum coherence in photosynthetic systems \cite{ish09}. A recent experiment with  photons in a controllable non-Markovian environment has been reported in Ref.~\cite{liu11}. Interestingly, we will show that non-Markovian dynamics can lead to smaller  quantum speed limit times.

\textit{Geometric approach.} We consider a possibly driven  open quantum system and  ask for the minimal time  that is necessary for it to evolve from an initial  state $\rho_0$  to a final state $\rho_\tau$. Without loss of generality, we assume that the initial state is pure, $\rho_0=\ket{\psi_0}\bra{\psi_0}$. The case of an initially mixed state can be treated by purification in a sufficiently enlarged Hilbert space \cite{joz94}.
 Note that under  nonunitary dynamics,  the final state $\rho_\tau$ will be generally   mixed. The basis of our geometric approach is   provided by the Bures angle $\mc{L}(\rho_0,\rho_\tau)$  between initial and final states of the quantum system \cite{bur69,joz94}, 
\begin{equation}
\label{eq02}
\mc{L}(\rho_0,\rho_\tau)=\arccos\left(\sqrt{\bra{\psi_0}\rho_\tau\ket{\psi_0}}\right).
\end{equation}
The Bures angle is a generalization to mixed states of the angle in Hilbert space between two state vectors \cite{nie00}.

To evaluate the quantum speed limit time, we consider the dynamical velocity  with   which the density operator of the  system evolves \cite{ana90}. The latter is given by the time derivative of the geometric Bures angle \eqref{eq02}, \begin{equation}
\label{eq03}
\begin{split}
\frac{\td}{\td t}\mc{L}(\rho_0,\rho_t)&\leq\left|\frac{\td}{\td t}\mc{L}(\rho_0,\rho_t) \right|\\
&=\frac{1}{\sqrt{1-\bra{\psi_0}\rho_\tau\ket{\psi_0}}}\,\frac{\left|\bra{\psi_0}\dot\rho_t\ket{\psi_0}\right|}{2\sqrt{\bra{\psi_0}\rho_\tau\ket{\psi_0}}}.
\end{split}
\end{equation}
Using the definition  \eqref{eq02}, Eq.~\eqref{eq03} can  be written as
\begin{equation}
\label{eq04}
2\co{\mc{L}}\si{\mc{L}}\,\dot{\mc{L}}\leq \left|\bra{\psi_0}\dot\rho_t\ket{\psi_0}\right|.
\end{equation}
Expression \eqref{eq04} will serve as the starting point for our unified derivation of  ML and MT type bounds on the rate of quantum evolution, using respectively the von Neumann trace inequality and the Cauchy-Schwarz inequality. 

\textit{Margolus-Levitin bound.} To illustrate the use of the von Neumann trace inequality, we begin by providing a  derivation of the ML bound  in the case of driven unitary dynamics. The generator is here given by the time-dependent Hamiltonian  $H_t$ of the system, and its density operator   obeys the von Neumann equation,
\begin{equation}
\label{eq05}
\dot\rho_t=\frac{1}{i\hbar}\,\left[H_t,\rho_t\right].
\end{equation}
Substituting Eq.~\eqref{eq05} into Eq.~\eqref{eq04}, we have,
\begin{equation}
\label{eq06}
2\co{\mc{L}}\si{\mc{L}}\,\dot{\mc{L}}\leq \frac{1}{\hbar}\left|\bra{\psi_0}\left[H_t,\rho_t\right]\ket{\psi_0}\right|.
\end{equation}
Equation \eqref{eq06} can be estimated from above with the help of the triangle inequality and we obtain,
\begin{equation}
\label{eq07}
2\co{\mc{L}}\si{\mc{L}}\,\dot{\mc{L}}\leq \frac{1}{\hbar}\left(\left|\tr{H_t\rho_t\,\rho_0}\right|+\left|\tr{\rho_t H_t\,\rho_0}\right|\right).
\end{equation}
To proceed, we introduce  the von Neumann  trace inequality for operators  which reads \cite{neu37,mir75,gri91},
\begin{equation}
\label{eq08}
\left|\tr{A_1 A_2} \right|\leq \sum\limits_{i=1}^n \sigma_{1,i} \sigma_{2,i}.
\end{equation}
Inequality \eqref{eq08} holds for any complex $n\times n$ matrices $A_1$ and $A_2$ with descending singular values, $\sigma_{1,1}\geq...\geq\sigma_{1,n}$ and $\sigma_{2,1}\geq...\geq\sigma_{2,n}$. The singular values of an operator $A$  are defined as the eigenvalues of $\sqrt{A^\dagger A}$ \cite{bha97}.  For a Hermitian operator, they are given by the absolute value of the eigenvalues of $A$, and are positive real numbers. If  $A_1$ and $A_2$ are simple (positive) functions of density operators acting on the same Hilbert space, Eq.~\eqref{eq08} remains true for arbitrary dimensions  \cite{gri91}. The singular values of the operators $A$ and $A^\dagger$ are moreover identical. By taking $A=(H_t\rho_t)^\dagger=\rho_t^\dagger H_t^\dagger=\rho_t H_t$, and combining Eqs.~\eqref{eq07} and \eqref{eq08}, we thus find,
\begin{equation}
\label{eq09}
2\co{\mc{L}}\si{\mc{L}}\,\dot{\mc{L}}\leq\frac{2}{\hbar}\,\sum_i \sigma_i p_i=\frac{2}{\hbar}\,\sigma_1,
\end{equation}
where $\sigma_i$ are the singular values of $H_t\rho_t$ and  $p_i=\delta_{i,1}$ those of the initial pure state $\rho_0$.  For a Hermitian operator, the operator norm  is given by the largest singular value, $||A||_\text{op}=\sigma_1$, while the trace norm  is equal to their  sum, $||A||_\text{tr}=\sum_i \sigma_i$ \cite{bha97}.  As a consequence, 
\begin{equation}
\label{eq10}
2\co{\mc{L}}\si{\mc{L}}\,\dot{\mc{L}}\leq\frac{2}{\hbar}\, ||(H_t\rho_t)||_\text{op}\leq\frac{2}{\hbar}\, ||(H_t\rho_t)||_\text{tr}.
\end{equation}
We note, in addition, that the trace norm is given by,
\begin{equation}
\label{eq11}
||(H_t\rho_t)||_\text{tr}=\tr{\left|H_t\rho_t\right|}=\la H_t \ra,
\end{equation}
when the instantaneous eigenvalues of $H_t$ are all positive (the latter can always be realized by properly choosing the zero of energy \cite{mar98}). Integrating Eq.~\eqref{eq10} over time from $t=0$ to $t=\tau$, we  arrive at the inequality,
\begin{equation}
\label{eq12}
\tau\geq\frac{\hbar}{2\,E_\tau}\sin^2(\mc{L}(\rho,\rho_\tau)),
\end{equation}
with the time-averaged energy $E_\tau=(1/\tau)\int_0^\tau\td t\,\la H_t\ra$. Equation \eqref{eq12} is the ML bound for driven closed systems. 

The above  derivation can be easily extended to arbitrary time-dependent nonunitary equations of the form, 
\begin{equation}
\label{eq13}
\dot\rho_t=L_t(\rho_t)\,,
\end{equation}
with positive generator $L_t$ \cite{bre07}. The  latter are trace class (super-)operators in a complex Banach space and need generally not be Hermitian. However, for symmetric norms, such as the Schatten $p$-norm, $||L_t||_\text{p} = [\sum_i \lambda_i^p]^{1/p}$, that we here consider \cite{rem2},  $||L_t^\dagger|| =   ||L_t||$. As a result, all previously used definitions and inequalities remain valid \cite{sim05}. Substituting Eq.~\eqref{eq13} into Eq.~\eqref{eq04}, we then have,
\begin{equation}
\label{eq15}
2\co{\mc{L}}\si{\mc{L}}\,\dot{\mc{L}}\leq \left|\bra{\psi_0}L_t(\rho_t)\ket{\psi_0}\right|,
\end{equation}
which is the nonunitary generalization of Eq.~\eqref{eq06}. Noting that $\bra{\psi_0}L_t(\rho_t)\ket{\psi_0}=\tr{L_t(\rho_t)\,\rho_0}$ and  employing the von Neumann trace inequality \eqref{eq08}, we obtain,
\begin{equation}
\label{eq16}
2\co{\mc{L}}\si{\mc{L}}\,\dot{\mc{L}}\leq\sum_i \lambda_i p_i=\lambda_1,
\end{equation}
where $\lambda_i$ are the singular values of the operator $L_t(\rho_t)$. Equation \eqref{eq16} can again be estimated from above by the operator norm and the trace norm to yield,
  \begin{equation}
\label{eq17}
2\co{\mc{L}}\si{\mc{L}}\,\dot{\mc{L}}\leq ||(L_t(\rho_t))||_\mrm{op}\leq ||(L_t(\rho_t))||_\mrm{tr}.
\end{equation}
Integrating Eq.~\eqref{eq17} over time, we eventually find,
\begin{equation}
\label{eq18}
\tau\geq \ma{\frac{1}{\Lambda_\tau^\mrm{op}}, \frac{1}{\Lambda_\tau^\mrm{tr}}}\sin^2(\mc{L}(\rho,\rho_\tau)),\end{equation}
where we have defined $\Lambda_\tau^\mrm{op,tr}=(1/\tau)\int_0^\tau\td t \, ||L_t(\rho_t)||_\mrm{op,tr}$. Equation \eqref{eq18} provides a ML type bound on the rate of quantum evolution valid for  arbitrary positive driven open system dynamics.

\textit{Mandelstam-Tamm bound.} We next  derive a unified bound for the quantum speed limit time for open systems by generalizing the method presented in Ref.~\cite{cam13} based on the relative purity.
 To this end, we rewrite Eq.~\eqref{eq04} as
\begin{equation}
\label{eq19}
2\co{\mc{L}}\si{\mc{L}}\,\dot{\mc{L}}\leq\left|\tr{L_t(\rho_t)\,\rho_0}\right|.
\end{equation}
The latter can be estimated from above with the help of the Cauchy-Schwarz inequality for operators:
\begin{equation}
\label{eq20}
2\co{\mc{L}}\si{\mc{L}}\,\dot{\mc{L}}\leq\sqrt{\tr{{L_t(\rho_t)\,L_t(\rho_t)^\dagger}}\,\tr{\rho_0^2}}.
\end{equation}
Since  $\rho_0$ is a pure state, $\tr{\rho_0^2}=1$, and we obtain,
\begin{equation}
\label{eq21}
2\co{\mc{L}}\si{\mc{L}}\,\dot{\mc{L}}\leq\sqrt{\tr{{L_t(\rho_t)\,L_t(\rho_t)^\dagger}}}=||L_t(\rho_t)||_\mrm{hs},
\end{equation}
where $||A||_\mrm{hs}= \sqrt{\tr{A^\dagger A}}=\sqrt{\sum_i \sigma_i^2}$ is the Hilbert-Schmidt norm \cite{bha97}.  Integrating Eq.~\eqref{eq12} over time leads to the following MT type bound for nonunitary dynamics, 
\begin{equation}
\label{eq22}
\tau\geq \frac{1}{\Lambda_\tau^\mrm{hs}}\sin^2(\mc{L}(\rho,\rho_\tau)),
\end{equation}
where $\Lambda_\tau^\mrm{hs}=(1/\tau)\int_0^\tau\td t\,||L_t(\rho_t)||_\mrm{hs}$.  For unitary processes, $\Lambda_\tau^\mrm{hs}$ is equal to the time-averaged  variance of the energy. For initially pure states,  relative purity and  fidelity are identical,  and the bound \eqref{eq22} thus reduces to the one derived in  Ref.~\cite{cam13}, since $\sin^2x\geq |\cos x -1|$, $0\leq x\leq \pi/2$; they are, however, different for initially mixed states.
Combining Eqs.~\eqref{eq18} and \eqref{eq22}, we obtain, 
\begin{equation}
\label{eq23}
\tau_\mrm{QSL}=\ma{\frac{1}{\Lambda_\tau^\mrm{op}},\frac{1}{\Lambda_\tau^\mrm{tr}}, \frac{1}{\Lambda_\tau^\mrm{hs}}}\sin^2(\mc{L}(\rho,\rho_\tau)).
\end{equation}
Equation \eqref{eq23} provides a unified expression for  the quantum speed limit time for generic (positive)  open system dynamics. It represents a  general extension of the MT and ML result, $\tau_\mrm{QSL}= \mbox{max}\{\pi\hbar/(2\Delta E),\pi \hbar/(2E)\}$.
 
We may go a step further by noting that for trace class operators  the following inequality holds (see Ref.~\cite{sim05}, Theorem 1.16),
\begin{equation}
\label{eq24}
||A||_\mrm{op} \leq ||A||_\mrm{hs}\leq ||A||_\mrm{tr}.
\end{equation}
As a result, $1/\Lambda_\tau^\mrm{op}\geq 1/\Lambda_\tau^\mrm{hs}\geq 1/\Lambda_\tau^\mrm{tr}$, and we can therefore conclude that the ML type bound based on the operator norm of the nonunitary generator provides the sharpest bound on the quantum speed limit time. We  will show below that the bound can be attained and  is hence tight.

\textit{Non-Markovian effects.} We may use the  above results to  investigate the influence of non-Markovian dynamics on the quantum speed limit time. 
To this end, we  consider the exactly solvable damped Jaynes-Cummings model for a two-level system resonantly coupled to a leaky single mode cavity \cite{bre99,gar99}; the environment is supposed to be initially in a vacuum state. The nonunitary generator of the reduced dynamics of the system is  
\begin{equation}
\label{eq27}
\begin{split}
L_t(\rho_t) =\gamma_t\left(\sigma_{-}\rho_t\sigma_{+} -\frac{1}{2}\,\sigma_{+}\sigma_{-}\,\rho_t -\frac{1}{2}\, \rho_t\,\sigma_{+}\sigma_{-}\right),
\end{split}
\end{equation}
where $\sigma_{\pm}=\sigma_x\pm i\sigma_y$ are the Pauli  operators and $\gamma_t$ the time-dependent decay rate. By assuming that there is only one excitation in the combined atom-cavity system, the environment can be described by an effective Lorentzian spectral density of the form,
\begin{equation}
\label{eq33}
J(\omega)=\frac{1}{2\pi}\frac{\gamma_0\lambda}{(\omega_0-\omega)^2+\lambda^2},
\end{equation}
where $\omega_0$ denotes the frequency of the two-level system, $\lambda$ the spectral width and $\gamma_0$ the coupling strength. The time-dependent decay rate is then explicitly given by,
\begin{equation}
\label{eq34}
\gamma_t = \frac{2\gamma_0 \lambda \sinh(dt/2)}{d\cosh(dt/2) + \lambda \sinh(dt/2)},
\end{equation}
where $d=\sqrt{\lambda^2-2\gamma_0\lambda}$. In the interaction picture, the reduced density operator  of the system at time $t$  reads,
\begin{equation}
\label{eq32}
\rho_t=\begin{pmatrix} \rho_{11}(0) \left|e^{-\int_0^t dt' \gamma_{t'}}\right| &\rho_{10}(0)e^{-\int_0^t dt' \gamma_{t'}/2}   \\ \rho_{10}^*(0) e^{-\int_0^t dt' \gamma_{t'}^*/2}  &1-\rho_{11}(0) \left|e^{-\int_0^t dt' \gamma_{t'}}\right|\end{pmatrix}.
\end{equation}
We shall examine the case where the system starts in the excited state, $\rho_{11}(0)=1$ and  $\rho_{10}(0) =0$.
For vanishing coupling, the system is isolated  and in a stationary state.  For finite coupling, the two-level system is driven by the bath.
The correlation time of the bath is $\tau_B = \lambda^{-1}$, while the decay time of the system is equal to  $\tau_S=\gamma_0^{-1}$. The non-Markovian properties of the model have been investigated in Refs.~\cite{wol08,xu10,lai10,hou11}: The dynamics is  Markovian  in the weak-coupling regime, $\gamma_0 < \lambda/2 $. For large time scale separation, $\tau_B\ll\tau_S$, or equivalently $\gamma_0 \ll \lambda$, the decay rate is constant, $\gamma_t = \gamma_0$.  The dynamics becomes non-Markovian for strong coupling, $\gamma_0>\lambda/2$, which corresponds to an imaginary parameter $d$. In this regime, the decay rate is an oscillatory  function of time. 
\begin{figure}
\includegraphics[width=0.48\textwidth]{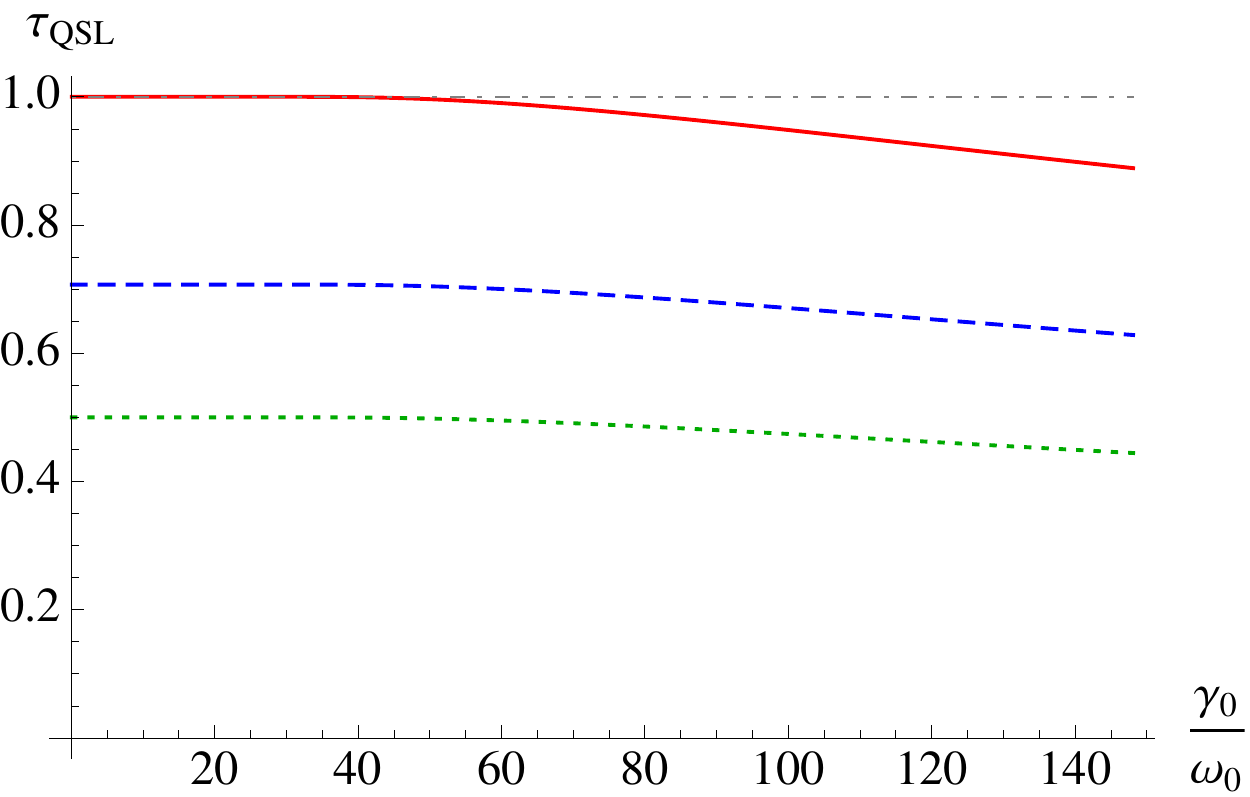}
\caption{\label{fig1}(color online) Quantum speed limit time $\tau_\mrm{QSL}$, Eq.~\eqref{eq23}, for the damped Jaynes-Cummings model  as a function of the coupling strength. The three curves are obtained for the trace norm (green dotted), Hilbert-Schmidt norm (blue  dashed) and the operator norm  (red solid) of the nonunitary generator $L(\rho_t)$, Eq. \eqref{eq27}. Parameters are $\lambda = 50$, $\omega_0=1$ and $\tau=1$.}
\end{figure}

\begin{figure}
\includegraphics[width=0.48\textwidth]{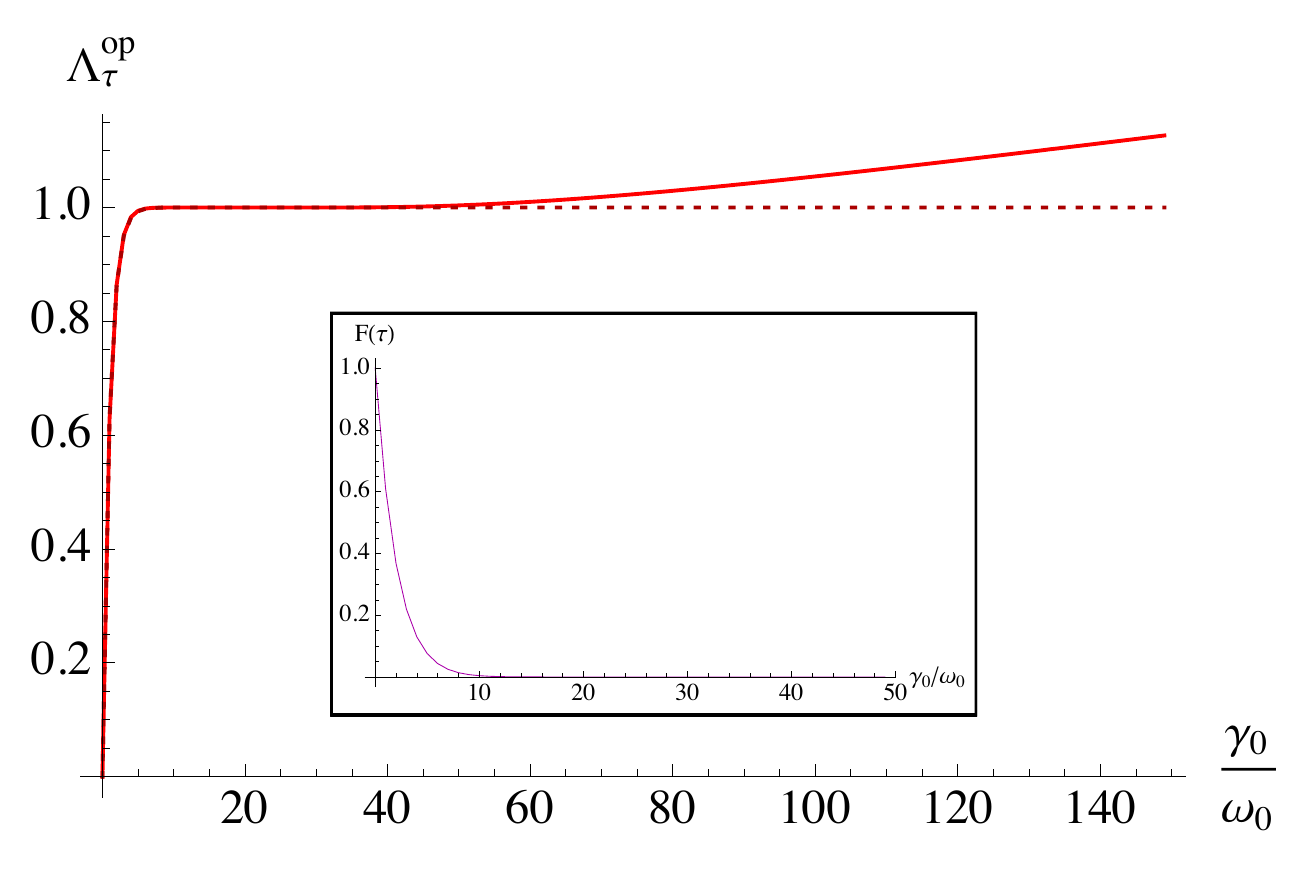}
\caption{\label{fig2}(color online) Time-averaged operator norm of the nonunitary generator $L(\rho_t)$ of the damped Jaynes-Cummings model \eqref{eq27}.  The dotted line corresponds to the Markovian result, Eq.~\eqref{33}. The inset shows the fidelity $F(\rho_0,\rho_\tau) = \cos  \mc{L}(\rho_0,\rho_\tau)$.  Same parameters as in Fig.~1.}
\end{figure}

Figure 1(a) shows the quantum speed limit time  \eqref{eq23} for the two-level system,  as a function of the coupling strength $\gamma_0$, obtained for the three different norms, in  the case $\tau =1$.  
We can distinguish two different phases. The speed limit time exhibits a plateau  independent of $\gamma_0$ for moderate coupling and then decreases for large coupling amplitudes.
Our second observation is  that the ML bound based on the operator norm is sharper than the MT bound based on the Hilbert-Schmidt norm, in agreement with Eq.~\eqref{eq24}. It is also sharper that the ML bound based on the trace norm. Remarkably, the operator-norm bound is \textit{tight} as it reaches the actual driving time $\tau$ over a large range of coupling strengths. 

The above behavior can be explained by evaluating the singular values of the operator $\dot \rho_\tau$ in the strong Markovian limit $\gamma_\tau=\gamma_0$. The two values are equal and given by $|\dot \rho_{11}|=|\dot \rho_{00}| = |\gamma_\tau \exp(-\int_0^{\tau} dt \gamma_{t})|= \gamma_0\exp(-\gamma_0 \tau)$.  For small coupling, $\gamma_0\tau\ll 1$,  the singular values are thus proportional to the coupling strength, $|\dot \rho_{11}|\simeq \gamma_0$. For larger coupling such that $\gamma_0 \tau\gg 1$, the singular values are independent of $\gamma_0$, $|\dot \rho_{11}|\simeq 0$. The plateau seen in Fig.~1 is hence a signature of Markovian dynamics and follows from the time independence of the decay rate. The height of the plateau can be determined by computing the time averaged norm of $\dot \rho_{t}$ plotted  in Fig. 2,
\begin{equation}
\label{33}
\frac{1}{\tau} \int_0^{\tau} dt\, ||\dot \rho_{t}||= \frac{n}{\tau}\left[1-\exp(-\gamma_0 \tau)\right],
\end{equation} 
with $n= 1, \sqrt{2} \mbox{ and } 2$ for the operator, Hilbert-Schmidt and trace norms, respectively. The constant $n$ is equal to $2^{1/p}$ for the general Schatten $p$-norm.  Equation \eqref{33} hence shows that the operator norm ($p= \infty$) is the only $p$-norm for which the  plateau reaches the actual driving time $\tau$.
Furthermore, the increase of the norm of the rate  $\dot \rho_{t}$ in the strong coupling regime, $\gamma_0>\lambda/2$,  appears as a consequence of  the (oscillatory) time dependence of the decay rate $\gamma_t$, and is thus a purely non-Markovian effect. We therefore reach the interesting conclusion that non-Markovian dynamics can increase the rate of evolution of a quantum system, and thus reduce the quantum speed limit time below its Markovian value. 

\textit{Conclusions.} We have derived a quantum speed limit time that generalizes the familiar MT and ML results to  generic time-dependent (positive) dynamics of open quantum systems. In particular, using the von Neumann trace inequality, we have obtained an expression of the speed limit time in terms of the operator norm of the nonunitary generator of the evolution. We have demonstrated that the latter bound is sharper than any bound based on a Schatten $p$-norm, such as the trace and  Hilbert-Schmidt norms, and that it is moreover tight. Applying these results to the damped Jaynes-Cummings model has additionally  shown that non-Markovian effects can lead to faster quantum evolution, and hence to  smaller quantum speed limit times.

This work was supported by the DFG (contract No LU1382/4-1). SD  acknowledges financial support by a fellowship within the postdoc-program of the German Academic Exchange Service (DAAD, contract No D/11/40955).

\end{document}